\documentclass[12pt]{iopart}

\usepackage{iopams}
\usepackage{epsfig,psfrag}
\usepackage{latexsym}
\begin{document}

\title[Fluctuation relations in Ising models]{Fluctuation relations in
non-equilibrium stationary states of Ising models.}

\author{A Piscitelli$^1$, F Corberi$^2$,
G Gonnella$^1$ and A Pelizzola$^3$}

\address{$^1$
Dipartimento di Fisica, Universit\`a di Bari {\rm and} Istituto Nazionale di Fisica Nucleare,
Sezione di  Bari, via Amendola 173, 70126 Bari, Italy}
\address{$^2$ Dipartimento di Matematica ed Informatica,
via Ponte don Melillo, Universit\`a di Salerno, 84084 Fisciano (SA), Italy }
\address{$^3$Dipartimento di Fisica {\rm and} Istituto
Nazionale di Fisica Nucleare, Sezione di  Torino, {\rm and} CNISM, Politecnico di Torino,
  c. Duca degli Abruzzi 24, 10129 Torino, Italy
}

\ead{\mailto{piscitelli@ba.infn.it}, \mailto{corberi@sa.infn.it},
\mailto{gonnella@ba.infn.it}, \mailto{alessandro.pelizzola@polito.it}}

\begin{abstract}

Fluctuation relations for the entropy production in non equilibrium stationary states of Ising
models  are investigated by Monte Carlo simulations. Systems  in contact with  heat baths at
two different temperatures or subject to external driving will be studied. In the first case,
by considering different kinetic rules and couplings with the baths,  the behavior  of the
probability distributions of the heat exchanged in a time $\tau$  with the thermostats, both
in the disordered and in the low temperature phase, are discussed. The fluctuation relation is
always verified in the large $\tau $ limit  and deviations from linear response theory are
 observed. Finite-$\tau$ corrections are  shown to obey a scaling behavior. In the
other case the system is in contact with a single heat bath but work is done by shearing it.
Also for this system the statistics collected for the mechanical work shows the validity of
the fluctuation relation and preasymptotic corrections behave analogously to the case  with
two baths.

\end{abstract}

\pacs{05.70.Ln; 05.40.-a; 75.40.Gb}

\maketitle

\section{Introduction} \label{intro}

In equilibrium statistical mechanics the knowledge of general expressions for the
probabilities of microscopic configurations is the cornerstone of a successful
 theory describing the macroscopic behavior of a large variety of
physical systems. Similar expressions, however, are not known for systems in non-equilibrium
steady states (NESS), despite their widespread occurrence in nature and their practical
interest.

NESS are usually realized by driving a system, either mechanically, as in the case of sheared
or stirred fluids, or thermodynamically, due for instance to couplings to reservoirs at
different temperatures. These states are characterized by a finite rate of entropy production,
and  the recent proposal~\cite{ECM93,ES94,GC95}  of a relation governing the fluctuations of
this quantity constitutes an important  result of general validity. The relation connects the
probability  $\mathcal{P}(\Sigma(\tau))$ of producing  an entropy $\Sigma(\tau)$ in a time
interval $\tau$, with the probability of  the opposite quantity, according to
\begin{equation}
 \ln{\frac{\mathcal{P}(\Sigma(\tau))} {\mathcal{P}(-\Sigma(\tau))}} =
- \Sigma(\tau) . \label{FR}
\end{equation}
Eq.~(\ref{FR}), also known as Gallavotti-Cohen relation, holds in the large $\tau$ limit,
specifically with $\tau$ larger than all relaxation times of the system. It was proved as a
theorem for a specific class of dynamical systems in \cite{GC95} and then established for
stochastic kinetics in \cite{Kurchan,LS,maes}. Expressions related to Eq.~(\ref{FR}) have been
established in Refs. \cite{Jarzynski,Crooks,HatanoSasa,VZC,vZCC,seifert,gawedzki}. Recent
reviews are given in \cite{rev}.

Fluctuation relations (FRs) are expected to be relevant in mesoscopic systems, particularly in
nano-- and biological sciences \cite{Bustamante}, because at these scales typical thermal
fluctuations may be sufficiently large to be comparable with the magnitude of the external
driving. FRs have nowadays been tested in some experiments \cite{wang,cili1,cili2}.

In this work  we study the FRs in simple standard statistical models, where their validity can
be ascertained, and some of the mechanisms of their occurrence can be investigated. We
consider the Ising model as a paradigmatic example of interacting system interested by a phase
transition. The model is maintained in NESS either by coupling it to two thermal baths at
different temperatures $T_1,T_2$, or by a mechanical forcing. In the former case, introducing
$\Delta \beta ^{(1)}= -\Delta \beta ^{(2)}=\left(\frac{1}{T_{2}} - \frac {1}{T_1} \right)$,
the relation (\ref{FR}) can be specified as
\begin{equation}
 \ln{\frac{\mathcal{P}(\mathcal{Q}^{(n)}(\tau))} {\mathcal{P}(-\mathcal{Q}^{(n)}(\tau))}}
  =
\mathcal{Q}^{(n)}(\tau) \Delta \beta^{(n)} , \label{FR2T}
\end{equation}
where $Q^{(n)}({\tau})$ is the heat exchanged with the heat bath at temperature $T_n$
($n=1,2$) in a time $\tau $. With mechanical driving Eq.~(\ref{FR}) can be cast as
\begin{equation}
 \ln{\frac{\mathcal{P}(\mathcal W(\tau))} {\mathcal{P}(-\mathcal W(\tau))}} =
 \frac{\mathcal{W}(\tau)}{T}, \label{FRW}
\end{equation}
where $\mathcal{W}(\tau) $ is the work done on the system  in the time interval $\tau$ and $T$
is the temperature of the thermostat.

In the case of thermodynamic driving,
the relation  (\ref{FR2T}) has been previously studied for different systems.
 It was shown to hold~\cite{lepri} for a chain of
oscillators coupled at the extremities to two thermostats, and it was studied~\cite{visco} for
a Brownian particle in contact with two reservoirs. A fluctuation relation has been proved
in~\cite{JW07} for the heat exchanged between two systems initially prepared in equilibrium at
different temperatures and later put in contact. The case of the Ising model coupled to
different reservoirs, similar to that considered in this paper,  has been  studied
analytically in~\cite{LRW}, in a mean-field approximation. Fluctuation properties of work due
to a magnetic field in transient regimes of Ising model have been analyzed in \cite{MEX}. We
are not aware of studies of the relation (\ref{FRW}) in stationary states of mechanically
driven Ising models.

In this Article we show that the FRs hold in the large--$\tau$ limit for the nearest neighbour
Ising model and both kinds  of  NESS analyzed. In the case of NESS induced by the presence of
two thermostats, we  also consider  different couplings with the baths and kinetic rules, some
of which have been previously reported in \cite{noi}, in order to address the generality of
the results. Our numerical data allow to appreciate and characterize the finite time
corrections to the asymptotic result. These were shown to be of order $1/\tau$ in
\cite{GC95,MR03,Giuliani05}. For systems described by a Langevin equation, in cases
corresponding to the experimental setup consisting of a resistor and a capacitor in parallel,
finite time corrections also behave as $1/\tau$ when work fluctuations are considered
\cite{vZCC,cili1}, while  faster decays  have been predicted for other topologies of circuits
\cite{vZCC}. Our data show that the leading term of such corrections decays as $1/\tau$. We
also propose an expression which well describes the corrections in an extended range of values
of $\tau$, incorporating the sub-leading behavior. This expression implies a scaling behavior
which takes into account the geometry of the system and the nature of the coupling with the
heat baths.

The occurrence  of a phase transition in the Ising model allows us to discuss the interplay
between the breaking of ergodicity and the validity of the fluctuation relation. A finite size
system in the low temperature NESS remains trapped into broken symmetry states for a time
$\tau _{erg}$ that diverges in the thermodynamic limit, much like in equilibrium. By varying
the system size and the baths temperatures, we are able to investigate the large-$\tau $ limit
both in the regime $\tau \gg \tau _{erg}$ and  $\tau \ll \tau _{erg}$. The latter is
particularly interesting because in this case, since FRs are expected to hold for $\tau $ much
larger than the characteristic relaxation times of the systems, they are not necessarily
obeyed in this condition. Interestingly, instead, we find that, the FR (\ref{FR2T}) holds true
in any case.

Close to equilibrium the FR implies the Green-Kubo relation (GKR)~\cite{galla96}. Therefore,
as discussed in~\cite{Giuliani05} a stringent test of the FR, which cannot be reduced to
linear response theory, can  be only achieved when the drive is large enough to bring the
system far from equilibrium, spoiling the GKR and/or determining non-Gaussian
$\mathcal{P}(\mathcal{Q}^{(n)})$. An analysis of the data in  the case of thermodynamic
driving will allow us to provide a
 strict test of the validity of the FR also in the far from equilibrium regime.

The paper is organized as follows. In the next section we introduce the model and discuss two
different implementations of the coupling to the heat baths. Then the results of our
simulations  will be presented and interpreted in terms of scaling expressions for finite time
corrections.  Relations with linear response theory are discussed in Sect. 2.3. In Section 3
the model with mechanical driving is considered.
 Section 4 includes our conclusions and a discussion of the perspectives of this work.

\section{Systems in contact with heat baths at two temperatures}

\subsection{The models}

We consider a two--dimensional Ising model defined by the Hamiltonian $ H\{\sigma \} = - J \sum_{\langle
ij \rangle} \sigma_i \sigma_j$, where $\sigma_i = \pm 1$ is a spin variable on a site $i$ of a
rectangular lattice  with $N=M \times L$ sites, $ \{\sigma \}$ is the configuration of all the spins
and the sum is over all pairs $\langle ij
\rangle $ of nearest neighbors.

In the case of {\it statical} coupling with the heat baths the system is  divided into two
halves. The left part (the first $M/2$ vertical lines) interacts with the heat bath at
temperature $T_1$ while the right part is in contact with the reservoir at $T_2>T_1$. We have
implemented both open or periodic boundary conditions. We used Monte Carlo spin-exchange
(Kawasaki) dynamics, corresponding to systems  with conserved magnetization. The case with
single spin dynamics was considered previously \cite{noi}. In the present case with two heat
baths, we have implemented the Kawasaki rule as follows: picking at random a couple of nearest
neighbor spins $\sigma _l,\sigma _m$ we attempt their exchange according to standard
Metropolis transition rates
\begin{equation}
{\mathcal A}(\{\sigma'  \}, \{\sigma \}) =min \left
\{\exp \left [-\frac{\Delta E (\{\sigma \},\{\sigma'  \})}{T}\right ],1\right \},
\label{metropolis}
\end{equation}
where $ \{\sigma \}$ and $\{\sigma'  \}$ are the configurations before and after the move and
$\Delta E (\{\sigma \},\{\sigma'  \})= H\{\sigma ' \}-H\{\sigma \}$. The temperature $T$ to be
entered in  ${\mathcal A}(\{\sigma'  \}, \{\sigma \}) $ is chosen as follows: if both the
spins considered are in contact with the same bath, $T$ is the temperature of that thermostat.
In the case in which, say, $\sigma_m$ is coupled to the temperature $T_1$ and $\sigma_l$ to
$T_2$,
 we compute $\Delta E (\{\sigma \},\{\sigma'  \})/T$ by splitting
the two contributions from the different reservoirs, namely $\Delta E (\{\sigma \},\{\sigma'
\})/T= - (J/T_2) \sigma '_l\sum_{ \langle i \rangle_l} \sigma '_i - (J/T_1) \sigma '_m\sum_{
\langle i \rangle_m} \sigma '_i +(J/T_2) \sigma _l\sum_{ \langle i \rangle_l} \sigma_i +
(J/T_1) \sigma _m\sum_{ \langle i \rangle_m} \sigma_i$, where $\langle i \rangle _l$ ($\langle
i \rangle _m$) are nearest neighbors of $\sigma_l$ ($\sigma _m)$.

In the second implementation, using single--spin dynamics, each spin $\sigma _i$, at a given
time $t$, is put {\it dynamically } in contact with one or the other reservoir depending on
the (time dependent) value of $h_i=(1/2)\vert \sum _{\langle j\rangle _i}\sigma _j \vert$,
where the sum runs over the nearest neighbor spins $\sigma _j$ of $\sigma _i$. Notice that
$h_i$ is one half of the (absolute value) of the local field. In two dimensions, with periodic
boundary conditions, the possible values of $h_i$ are $h_i=0,1,2$. At each time, spins with
$h_i=2$ are connected to the bath at $T=T_1$ and those with $h_i=1$ with the reservoir at
$T=T_2$. Namely, when a particular spin $\sigma _i$ is updated, the temperature $T_1$ or $T_2$
is entered into the transition rate according to the value of $h_i$. Loose spins with $h_i=0$
can flip back and forth regardless of temperature because these moves do not change the energy
of the system. Then, as in the usual Ising model, they are associated to a temperature
independent transition rate. Notice that $h_i=2$ correspond to spins whose surrounding
neighbors are aligned, a situation which is typically found in the bulk of ordered domains,
while $h_i=1$ corresponds to interfacial spins. This model was introduced in~\cite{olivera}
and studied in~\cite{godreche}. It is characterized by a line of critical points in the plane
$T_1,T_2$, separating a ferromagnetic from a paramagnetic phase analogously to the equilibrium
Ising model.  Metropolis transition rates  have  been considered also in this case.

Considering the stationary state, a generic evolution of the system is given by the sequence
of configurations $\{\sigma (t)\}=\{\sigma _1(t), \dots , \sigma _N (t)\}$ where
$\sigma _i(t)$ is the value
of the spin variable at time $t$. Denoting with $t^{(n)}_k $ the times (measured here as the
number of elementary Montecarlo updates) at which an elementary move is attempted by coupling
the system to the $n$-th reservoir, the heat released by the bath in a time window
$[s,s+\tau]$ is defined as
\begin{equation}
\mathcal{Q}^{(n)}(\tau) = \sum_{\{t^{(n)}_k \} \subset [s, s+\tau]} [H(\sigma(t^{(n)}_k)) -
H(\sigma(t^{(n)}_k-1))]. \label{calore}
\end{equation}
In the case of dynamic coupling to the thermostats, recalling the discussion above,
$\mathcal{Q}^{(1)}(\tau)$ and $\mathcal{Q}^{(2)}(\tau)$ will be also referred to as {\it bulk}
and {\it interface} exchanged heats. The properties of $\mathcal{Q}^{(n)}(\tau)$ will be
computed by collecting the statistics over different sub-trajectories obtained by dividing a
long history of length $t_F$ into many ($t_F/\tau$) time-windows of length $\tau $, starting
from different $s$.

Notice that all the dynamical rules considered insofar obey the generalized detailed balance
condition \cite{BD}
\begin{equation}
e^{\mathcal{Q}^{(1)}/T_1 + \mathcal{Q}^{(2)}/T_2}
\mathcal{A}(\{\sigma'\},\{\sigma\}) =
\mathcal{A} (\{\sigma\},\{\sigma'\})
\label{GDB}
\end{equation}
where $\mathcal{Q}^{(1)}, \mathcal{Q}^{(2)}$  are the heats  exchanged with the reservoirs
during an elementary  transition.

\subsection{Results for $ {T_1,T_2>T_c}$}

 We begin our analysis with the study  of the relation~(\ref{FR2T}) in
the case with  both temperatures above the critical value $T_c\simeq 2.269$ of the equilibrium
Ising model. In the following we  will  measure times in montecarlo steps (MCS) (1 MCS=N
elementary moves) and set $J=1$.

\begin{figure}
\vskip 1cm
\includegraphics*[height=6.5cm,width=11.cm]{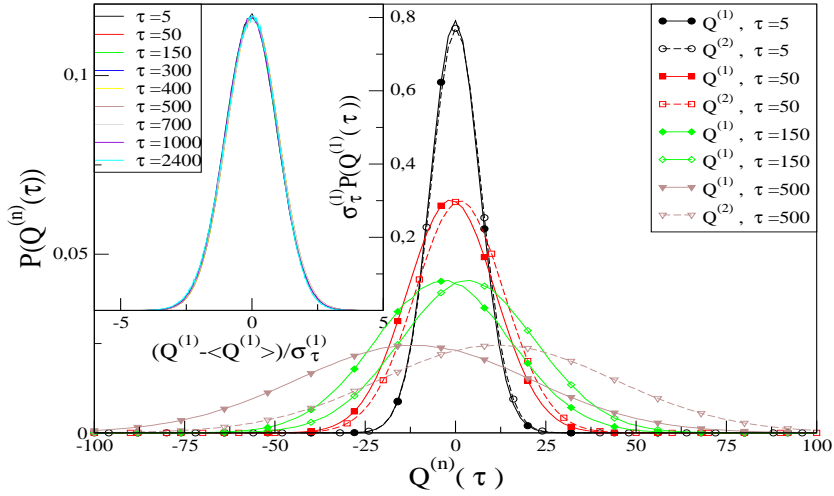} \vskip 0.4cm
\vskip 0.5cm
 \caption{ Heat PDs for $\mathcal{Q}^{(1)}$ (on the left) and
 $\mathcal{Q}^{(2)}$ (on the right) for a system evolving
with spin--exchange--dynamics
 and $T_1 =2.9, T_2=3$, size 20x20 and
$t_F= 5 \cdot 10^8 $  MCS. In the inset
$\sigma^{(1)}_{\tau}\mathcal{P}(\mathcal{Q}^{(1)}({\tau}))$ is plotted against
$(\mathcal{Q}^{(1)}(\tau) - \langle \mathcal{Q}^{(1)}(\tau)\rangle)/\sigma^{(1)}_{\tau}$.
Curves for different $\tau $ collapse on a Gaussian mastercurve. The same collapse of data
would be observed for  $\mathcal{Q}^{(2)}$.} \label{pdf_kawas_aboveTc}
\end{figure}

\begin{figure}
\includegraphics*[height=6.5cm,width=9.cm]{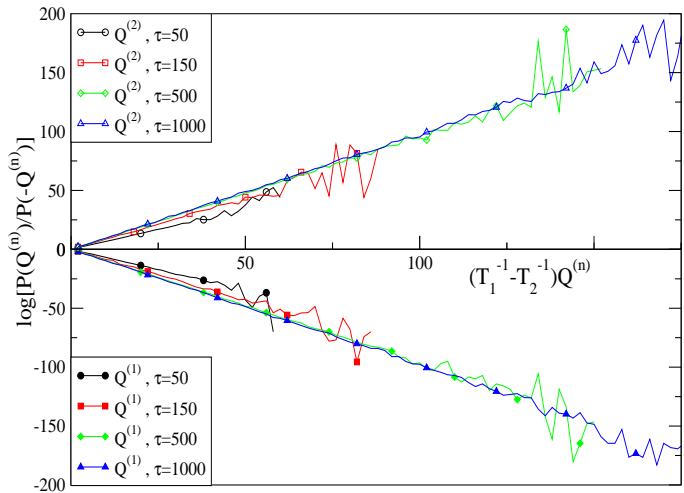} \vskip 0.4cm
\vskip 1cm
 \caption{Same parameters of Fig.~\ref{pdf_kawas_aboveTc}.
  $\log \left [\mathcal{P}(\mathcal{Q}^{(n)}(\tau))/ \mathcal{P}(-\mathcal{Q}^{(n)}(\tau))\right
]$ is plotted against $(1/T_1-1/T_2)\mathcal{Q}^{(n)}(\tau)$ ($n=1$ lower panel, $n=2$ upper panel).}
\label{rette_kawa_aboveTc}
\end{figure}

\begin{figure}
\includegraphics*[height=6.5cm,width=9.cm]{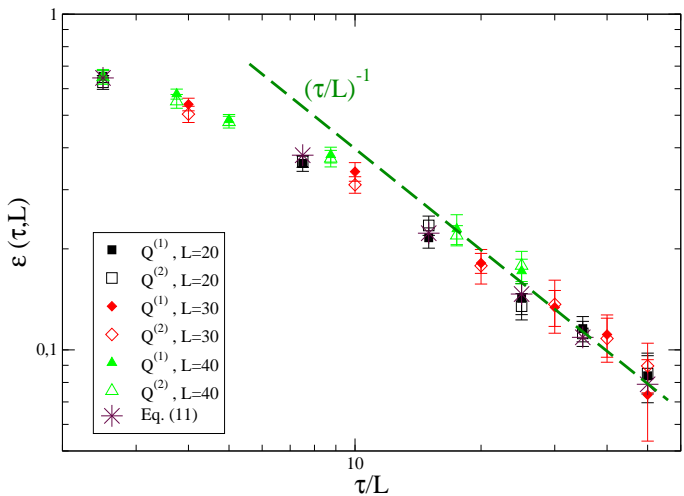} \vskip 0.4cm
\vskip 1cm
 \caption{Same parameters of Fig.~\ref{pdf_kawas_aboveTc} (different sizes).
  $\epsilon (\tau,L)$ is plotted against $\tau /L$ for different $L$.
Results from a fit based on  formula (\ref{deviazione1}) are also shown. }
\label{scala_kawa_aboveTc}
\end{figure}

The typical behavior of the heat probability distributions (PD)  is reported in
Fig.~\ref{pdf_kawas_aboveTc} for the system  with static coupling to the baths, $T_1=2.9$,
$T_2=3$, and a square geometry with $L=M=20$.  Much larger sizes can be hardly used because
trajectories with a heat whose sign is opposite to that of the average value would be too
rare. Results are qualitatively similar to those obtained with non-conserved dynamics
\cite{noi}.
 $\mathcal{Q}^{(1)}(\tau)$ ($\mathcal{Q}^{(2)}(\tau)$) is on average
negative (positive) and the relation
\begin{equation}
\langle \mathcal{Q}^{(1)}(\tau)\rangle + \langle \mathcal{Q}^{(2)}(\tau)\rangle =0
\label{avenull}
\end{equation}
is verified. The PDs for $\mathcal{Q}^{(1)}$ and $\mathcal{Q}^{(2)}$ are  similar but when
$\Delta T = T_2 - T_1$ is increased the distribution of $\mathcal{Q}^{(1)}$ can be observed to
be more peaked.

The only characteristic time
 above $T_c$ is the (microscopic) relaxation time that can be measured from the decay
 of the autocorrelation function; it is of few  MCS  for the case of
Fig.~\ref{pdf_kawas_aboveTc}. At sufficiently large  $\tau$,  greater  than  the relaxation
time, due to the central limit theorem, one expects a Gaussian behavior for the probability
distributions, namely $\mathcal{P}(\mathcal{Q}^{(n)}({\tau}))= (2\pi
(\sigma^{(n)}_{\tau})^2)^{-1/2} \exp [- \frac {(\mathcal{Q}^{(n)}(\tau) - \langle
\mathcal{Q}^{(n)}(\tau)\rangle)^2} {2(\sigma^{(n)}_{\tau})^2}]$,  with $\langle
\mathcal{Q}^{(n)}_\tau\rangle \sim \tau$ and $\sigma^{(n)}_{\tau}\sim \sqrt \tau$. This form
is found with good accuracy, as shown by the collapse of the PD's at different times in the
inset of Fig.~\ref{pdf_kawas_aboveTc}.

In Fig.~\ref{rette_kawa_aboveTc}, in order to study the  FR (\ref{FR2T}),  the logarithm of
the ratio $\mathcal{P}(\mathcal{Q}^{(n)}(\tau))/ \mathcal{P}(-\mathcal{Q}^{(n)}(\tau))$ is
plotted as a function of $\Delta \beta^{(2)}\mathcal{Q}^{(n)}(\tau)$.
 For every value of $\tau $ the data are well consistent with a linear relationship even if,
 for large values of the heat,  the statistics becomes poor.
The  FR (\ref{FR2T}) is verified if  the slopes
\begin{equation}
D^{(n)}(\tau)= \frac{\ln{\frac{\mathcal{P}(\mathcal{Q}^{(n)}(\tau))}
{\mathcal{P}(-\mathcal{Q}^{(n)}(\tau))}}}{ \mathcal{Q}^{(n)}(\tau) \Delta \beta^{(n)}}
\label{slope}
\end{equation}
tend to 1 when $\tau \to \infty$. In Fig.~\ref{scala_kawa_aboveTc} the behavior of the {\it
distance} $\epsilon^{(n)}=1-D^{(n)}(\tau)$ from the expected asymptotic result  is shown for
the case of Fig.~\ref{pdf_kawas_aboveTc}.  Indeed, this quantity  goes to zero for large $\tau
$ showing the validity of Eq.~(\ref{FR2T}).

\begin{figure}
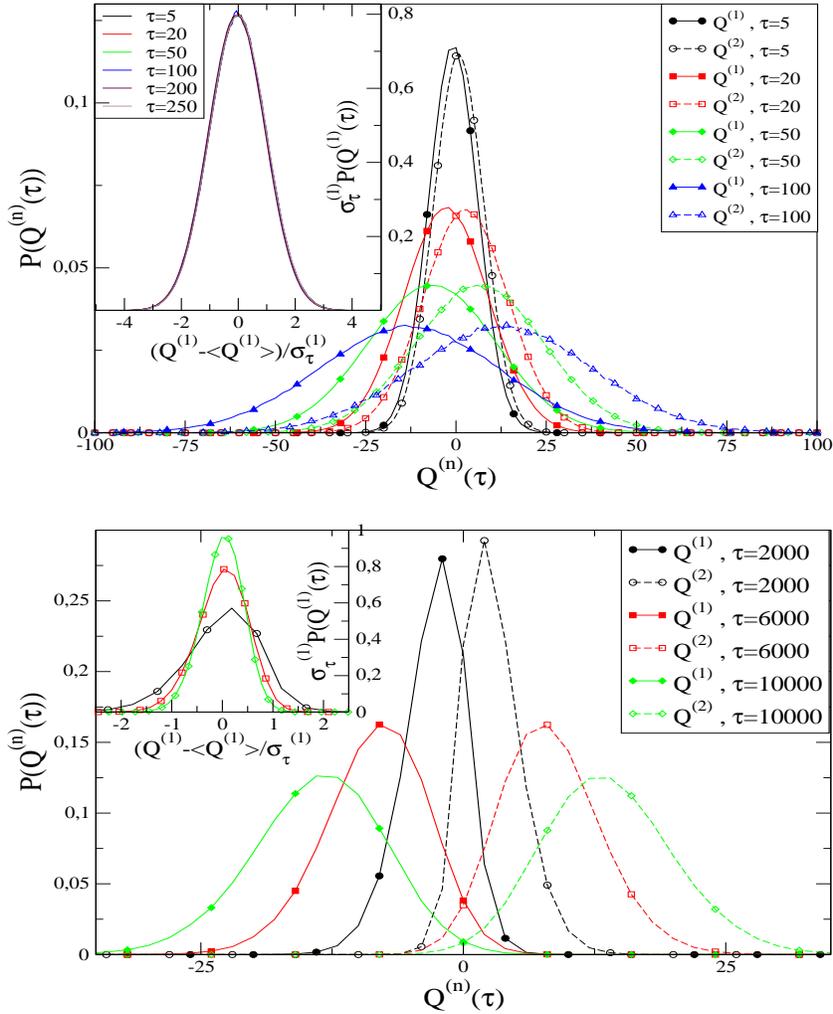

\includegraphics*[height=6.5cm,width=11.cm]{figlione4a.eps} \vskip 0.4cm
\includegraphics*[height=6.5cm,width=11.cm]{figlione4brr.eps}
\vskip 1cm \caption{Upper panel ($T_1,T_2>T_c$): Heat PDs for bulk heat $\mathcal{Q}^{(1)}$
(on the left) and interface heat
 $\mathcal{Q}^{(2)}$ (on the right) for the system with dynamical coupling
and $T_1 =3.0,  T_2=3.1$, size 10x10 and $t_F= 5 \cdot 10^8$ MCS. In the inset
$\sigma^{(1)}_{\tau}\mathcal{P}(\mathcal{Q}^{(1)}({\tau}))$ is plotted against
$(\mathcal{Q}^{(1)}(\tau) - \langle \mathcal{Q}^{(1)}(\tau)\rangle)/\sigma^{(1)}_{\tau}$.
Curves for different $\tau $ collapse on a Gaussian mastercurve. Lower panel ($T_1,T_2<T_c$):
Same kind of plot for $T_1 =1, T_2=1.3$, size 10 x 10 and $t_F= 10^{9} $ MCS.  The inset shows
that  PD's  at different $\tau$  do not collapse on a single curve after rescaling. }
 \label{pdfmendezaboveTc}
\end{figure}

A similar behavior occurs in the case with dynamic couplings. In the upper panel of
Fig.~\ref{pdfmendezaboveTc} the PD's for the interface and bulk exchanged heats
$\mathcal{Q}^{(2)}$ and $ \mathcal{Q}^{(1)}$ are shown. The PD's corresponding to the colder
bath  are always   higher and narrower. As in the case with static coupling, also now the PD's
at different $\tau$  can be rescaled on a single Gaussian master curve, as it can be seen in
the inset of the upper panel of Fig.~\ref{pdfmendezaboveTc}. The distances $\epsilon^{(n)}$
tend to zero, as shown in Fig.~\ref{scalingmendez}, and the fluctuation relation (\ref{FR2T})
is verified. A scaling argument predicting the behavior of $\epsilon^{(n)}$ will be presented
in Sec.~\ref{finitetau}.

\subsection{Deviations from the Green-Kubo relation}

As discussed in the introduction, once the validity of the FR has been ascertained in this
context, it is interesting to see if deviations from the Green-Kubo relation are observed, in
order to provide a rigid test of the FR not related to linear response theory. The GKR for the
current $\mathcal{J} = |<\mathcal{Q}^{(n)}(\tau)>|/\tau$ (which does not depend on $n$ due to
Eq.~(\ref{avenull})) reads
\begin{equation}
\lim_{\Delta T \rightarrow 0} \, \frac{\mathcal{J}}{\Delta T} =
\frac{\mathcal D}{2T^2_1}
\label{GK}
\end{equation}
where $\mathcal D$ is related to the fluctuations of the heat $\mathcal{Q}$  exchanged with
the bath in the equilibrium state at $T_1$ through
\begin{equation}
\mathcal D = \frac{<\mathcal{Q}(\tau)^2>}{\tau} .
\end{equation}

In order to check if the GKR~(\ref{GK}) is verified we proceeded as follows: i) from an
equilibrium simulation at $T_1=3$ we have extracted $\mathcal D$ and ii) by fixing $T_1$ and
varying $\Delta T$ in the range $[0,1]$ we have computed the current $\mathcal J$. We have
used the largest values of $\tau $ for which the slope (\ref{slope}) is measurable and the FR
was verified.

\begin{figure}
\includegraphics*[height=11cm,width=9.cm,angle=270]{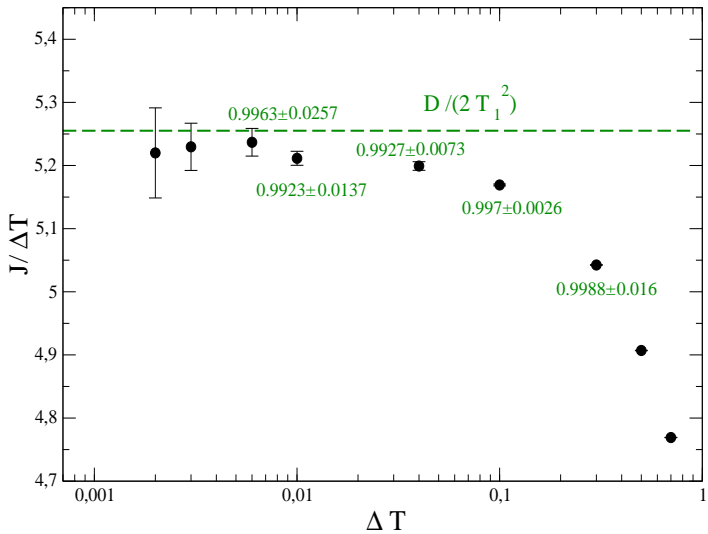}
\caption{The quantity $\mathcal J/\Delta T$ is plotted against $\Delta T$ for the system with
dynamic coupling to the baths, size 10x10,  $T_1=3$.
 The horizontal dashed
line is the value of $\mathcal D/(2T_1^2)$ obtained from the simulation of the system in the
equilibrium state at $T_1$. Near some of the points the values of the slopes $D^{(1)}(\tau)$
at $\tau=450$ have been reported.} \label{figgk}
\end{figure}

The data for the case with dynamical coupling to the baths are shown in
Fig.~\ref{figgk}.
Our simulations clearly show deviations from the GKR~(\ref{GK}) in
a range of
$\tau$ where the FR holds.
The same conclusion is arrived at in the case with static coupling to the
reservoirs. This proves that in the cases considered insofar the FR is verified beyond the
linear regime.

\subsection{Scaling behavior of finite-$\tau$ corrections} \label{finitetau}

Generally, $\epsilon^{(n)}$ depend on the temperatures, the geometry of the system and on
$\tau $. We propose now a scaling argument for the behavior of this quantity that takes also
in account the nature of the coupling with the reservoirs.
Considering a trajectory of length $\tau$ in configuration space, enforcing Eq.~(\ref{GDB}),
one can compute the ratio between the probability of the trajectory $\mathcal{P} (traj)$,
conditioned to a given initial state, and that of the time-reversed evolution, obtaining
\begin{equation}
\frac{\mathcal{P} (traj)}{\mathcal{P}(-traj)} = e^{-\frac {\mathcal{Q}^{(1)}(\tau)}{T_1} -
\frac {\mathcal{Q}^{(2)}(\tau)}{T_2}} =  e^{\mathcal{Q}^{(n)}(\tau) \Delta \beta^{(n)} -
\frac{\Delta E}{T_{n'}}}, \label{microrev}
\end{equation}
where $n'\ne n$ and  $\Delta E=\mathcal{Q}^{(1)}(\tau) + \mathcal{Q}^{(2)}(\tau)$ is the
difference between the energies of the final and initial states. From Eq.~(\ref{microrev}),
after averaging over all possible  trajectories \cite{BD}, it is  straightforward  to  arrive
to the following expression
\begin{equation}
\langle \frac{\mathcal{P}^{staz}(\tau)}{\mathcal{P}^{staz}(0)} e^{\frac{\Delta E}{T_{n'}}}
\rangle =\langle e^{-{\epsilon^{(n)}}{\Delta \beta^{(n')}} Q^{(n)}(\tau)} \rangle ,
\label{piscia}
\end{equation}
where ${\mathcal{P}^{staz}(0)}$ (${\mathcal{P}^{staz}(\tau)}$) is the probability of the initial (final) state.

In order to proceed further, we make a {\it quasi equilibrium} hypothesis, namely
\begin{equation}
\frac{\mathcal{P}^{staz}(\tau)}{\mathcal{P}^{staz}(0)} \simeq e^{-\beta_1 \Delta
E_1 - \beta_2 \Delta E_2 }
\label{appro}
\end{equation}
with $\Delta E = \Delta E_1 + \Delta E_2$. This is a strong assumption which is not expected
to hold true in general. However  it should be correct when the non-equilibrium drive is
small, and the system is such that the NESS approaches the equilibrium state in the limit of
small drive \footnote{The fulfillment of this condition is not obvious, particularly in
systems with ergodicity breaking. It is not verified in the low temperature phase of the
large-N model under shear flow \cite{largeN}, a model related (but with a {\it mean field}
character) to the Ising one considered in this paper, nor in some driven mean field models,
see \cite{drivenmean}.}. In the present case, therefore, it is expected to apply for small
$\Delta T$,  which is indeed the case of our simulations, in the high temperature disordered
phase. Furthermore, assuming that $\Delta E_n$ and $Q^{(n)}$ are Gaussian distributed (which
has been actually verified numerically for $T_1, T_2
> T_c$), one finally arrives at
\begin{equation}
{\epsilon^{(n)}}{\Delta \beta^{(n)}} \simeq - \frac{\langle Q^{(n)}(\tau)
\rangle}{(\sigma^{(n)}_{\tau})^2} \mp \sqrt{\left( \frac{\langle Q^{(n)}(\tau)
\rangle}{(\sigma^{(n)}_{\tau})^2}\right) ^{2}+\frac{v^2_{\Delta E} }
{(\sigma^{(n)}_{\tau})^2}(\Delta\beta^{(n)})^{2} }
 \label{deviazione1}
\end{equation}
where the signs $-,+$ are  for $n=1,2$, respectively,  and $v^2_{\Delta E}$ is a quantity
related to the variance of $\Delta E_n$ which, for large $\tau$, is expected not to depend
much on $\tau$.

\begin{figure}
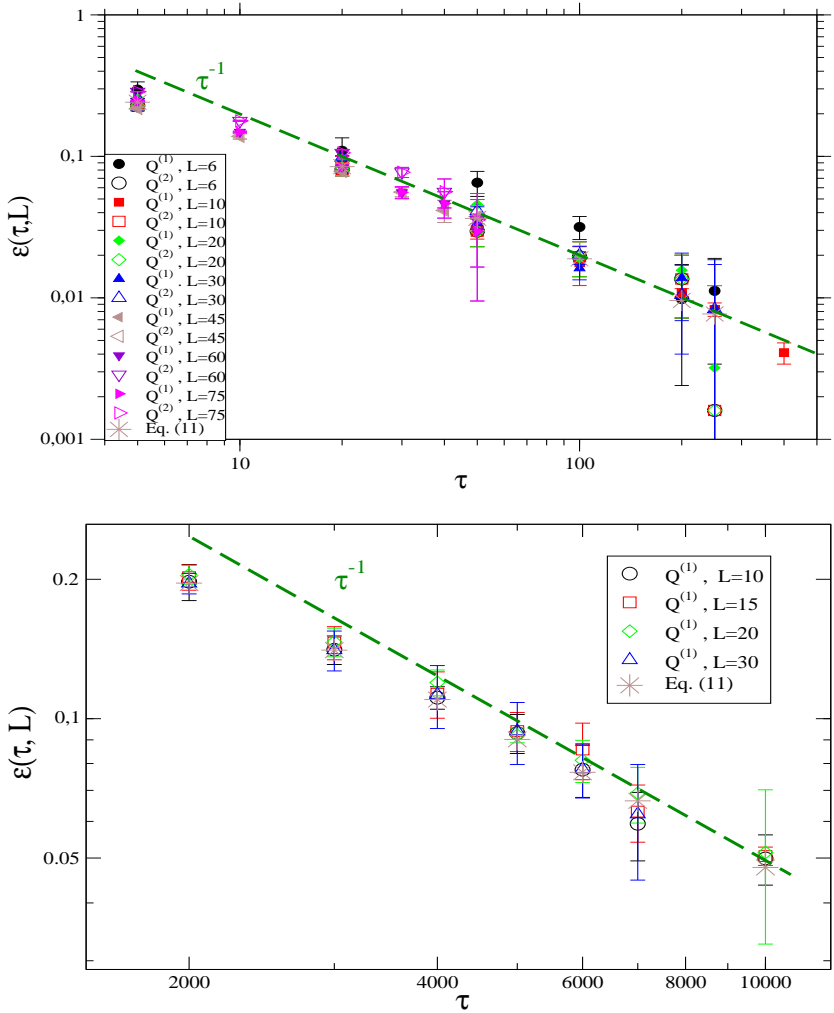

\includegraphics*[height=6.5cm,width=11.cm]{figlione5a.eps} \vskip 0.4cm
\includegraphics*[height=6.5cm,width=11.cm]{figlione5b.eps} \vskip 0.4cm
\vskip 1cm
 \caption{
Same parameters of Fig.~\ref{pdfmendezaboveTc} (different sizes). Upper Panel ($T_1,T_2>T_c$):
$\epsilon^{(n)} (\tau,L)$ are plotted against $\tau $ for different $L$.   Lower panel
($T_1,T_2<T_c$): $\epsilon^{(1)}$ is plotted against $\tau $ for different $L$. }
\label{scalingmendez}
\end{figure}

As discussed above, for large $\tau$,  the quantities $<Q^{(n)}(\tau)>$  and $(\sigma
^{(n)}_{\tau})^2$ grow proportionally to $\tau$, so that their ratio is asymptotically
constant. More precisely, since the heat is exchanged between neighboring spins coupled to
different reservoirs, indicating by  $N_{flux}$ the number of such couples of spins we expect
$\mathcal{Q}^{(n)}(\tau ) \propto (\sigma^{(n)}_{\tau})^2 \propto N_{flux} \tau$. In the model
with static coupling to the baths one has $N_{flux}\propto L$. With dynamic coupling, instead,
since every spin in the system can feel one or the other temperature, one has $N_{flux}\propto
N$. Since  $v^2_{ \Delta E}$ is an extensive quantity proportional to the number $N$ of spins,
from Eq.~(\ref{deviazione1}) one obtains the scaling form $\epsilon^{(n)}(\tau,L) \simeq
f(x)$, with
\begin{equation}
   x= \left \{ \begin{array}{ll}
        \tau L^{-1}  \qquad $static coupling to baths$  \\
        \tau  \qquad $dynamic coupling to baths$
        \end{array}
        \right .
        \label{scaling}
\end{equation}
and $f(x)$ is a (temperature dependent) scaling function with the large-$x$ behavior $f(x)\sim
x^{-1}$. Notice that the corrections to the universal law (\ref{FR2T}) are system dependent
but with the same asymptotic dependence on $\tau$. We will now examine the validity of this
scaling behavior in the cases considered so far.

For the system  with static coupling the data of Fig.~\ref{scala_kawa_aboveTc} confirm our
predictions: Curves with different $L$ collapse when plotted against $x=\tau /L$, for all the
values of $x$ considered, and $\epsilon ^{(n)}\propto x^{-1}$ for sufficiently large $\tau $.
We reported also the results obtained by applying formula (\ref{deviazione1}) with
$v^2_{\Delta E}$ used as a fitting parameter; we observe  a very good  overlap with the
numerical data. As expected, the fitting parameter results to be a quantity proportional to
the total number $N$ of lattice sites. Moreover, we considered different cases by varying
$\Delta \beta$ in the interval 0.1-0.4 at fixed $T_1$, and we did not find a significant
dependence of the fitting parameter  on $\Delta \beta$.

The forms (\ref{deviazione1}) and (\ref{scaling}) are   also verified for the Ising model with
dynamical coupling (see Fig.~\ref{scalingmendez}). Sizes between $L=6$ and $L=75$ have been
considered. The data collapse when plotted as a function of $\tau$ (except, perhaps, for  the
system  of size $L=6$ which is probably too small for our scaling argument to fully apply). At
large $\tau$ one observes $\epsilon^{(n)} \sim \tau^{-1}$, while preasymptotic results are
again well reproduced by Eq.~(\ref{deviazione1}).

\begin{figure}
\includegraphics*[height=9.5cm,width=11.cm]{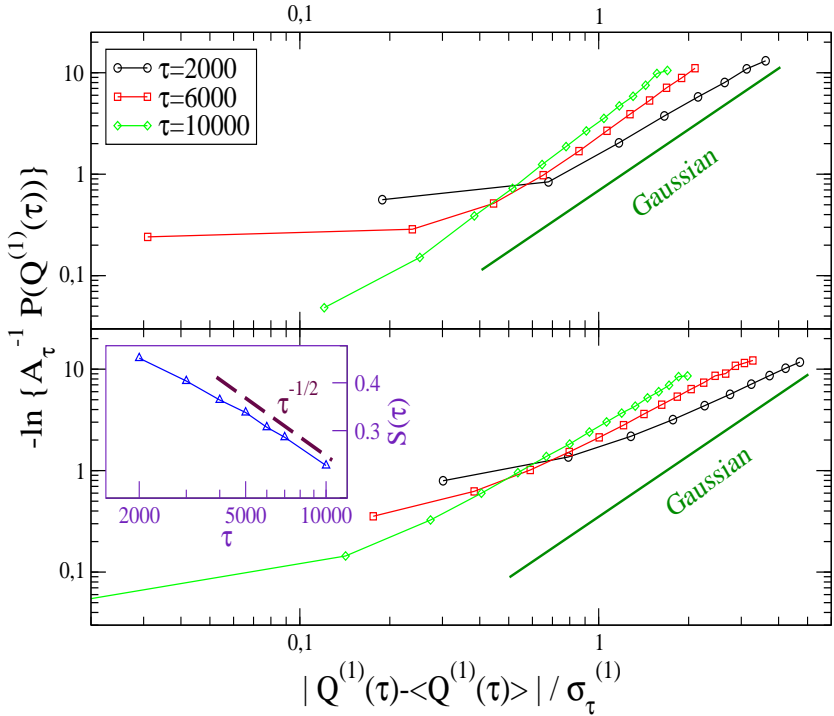}
\caption{The logarithm of the PDs of the lower panel of Fig.~\ref{pdfmendezaboveTc} (for
$Q^{(1)}$ only) (normalized by the height of $A_\tau$ of the peak) is plotted against
$|Q^{(1)}(\tau)-\langle Q^{(1)}(\tau)|/\sigma _\tau $. In the upper (lower) panel the branch
$Q^{(1)}< \langle Q^{(1)} \rangle $ ($Q^{(1)} > \langle Q^{(1)} \rangle$) is plotted. The
continuous line is the Gaussian behavior. In the inset the skewness is plotted against time.
The dashed line is the power law behavior $\tau ^{-1/2}$.}
 \label{pdfmendezaboveTcbis}
\end{figure}

\subsection {Results for $ {T_1,T_2 < T_c}$} \label {below}

Before discussing the results of the simulations in the low temperature phase it is useful to
overview the behavior of the Ising model in contact with a single bath at temperature $T<T_c$.
In the thermodynamic limit the system is confined into one of the two pure states, which can
be distinguished by the sign of the magnetization $m(T)=(1/N)\sum _{i=1}^N \sigma _i$. This
state is characterized by a finite relaxation time $\tau _{eq}(T)$. When $N$ is finite,
instead, genuine ergodicity breaking does not occur. Nevertheless, the system remains trapped
into the basin of attraction of the pure states also in this case, although only for a finite
time $\tau _{erg}(N,T)$. Therefore, for a finite size system in the low temperature phase,
there is the additional timescale $\tau _{erg}(N,T)$, beside $\tau _{eq}(T)$, which can grow
large. This ergodic time diverges when $N\to \infty$ or $T\to 0$, when $\tau _{eq}(T)$ is left
as the only time scale in the system. For what follows, it is important to recall that, due to
the trapping discussed above, an observation of a finite system on timescales much smaller
than $\tau _{erg}(N,T)$ is representative of the state with ergodicity breaking which,
however, strictly speaking, can be only realized for $N=\infty$.

A similar picture holds for   systems in contact with two thermal baths, where there are  two
{\it ergodic times}  $\tau _{erg}(N,T_n)$, one for each subsystem. Since the FR is expected to
hold for $\tau $ larger than the typical timescales of the system, it is interesting to study
the role of the additional timescales $\tau _{erg}(N,T_n)$ on the FR and the interplay between
$\tau $ and $\tau _{erg}(N,T_n)$.

We have studied numerically the model  with dynamic coupling to the baths at ${T_1,T_2 <
T_c}$. $\tau _{erg}(N,T_n)$ can be evaluated as the time over which the autocorrelation
functions $C^{(n)}(t-t')=\langle \sum _{i=1} ^N\sigma _i^{(n)} (t)\sigma _i^{(n)}(t')
\rangle$, where $\sigma _i^{(n)}$ denote spins in contact with the bath at $T=T_n$, decay to
zero (notice that $\tau _{erg}(N,T_1)>\tau _{erg}(N,T_2)$ since $T_1<T_2$). Then, by varying
$T_1,T_2$ and $N$ appropriately one can realize the limit of large $\tau $ in the two cases
with i) $\tau \ll \tau _{erg}(N,T_2)$ or ii) $\tau \gg \tau _{erg}(N,T_1)$. In the former
case, in the observation time-window $\tau $ the system is practically confined into pure
states while in the latter ergodicity is restored. Not surprisingly, in the latter case, we
have  observed a behavior very similar to that with $T_1,T_2>T_c$, except for the timescales
required to access the asymptotic stage which are much larger in the low temperature regime,
due to the tiny amount of heats exchanged. In particular, the PDs are found to be Gaussian.
More interesting is the case i), for which the PDs are shown in the lower panel of
Fig.~\ref{pdfmendezaboveTc}.  The distributions extend on a smaller support, compared with
those for  $T_1,T_2 > T_c$, even if the times over which the heats $\mathcal{Q}^{(n)}$ are
collected are larger (see insets of the upper panel of Fig.~\ref{pdfmendezaboveTc}). Moreover,
quite interestingly, in this case the PDs are not Gaussian in the range of times accessed by
the simulations. This can already be seen from the lower panel of Fig.~\ref{pdfmendezaboveTc},
since the PDs are not symmetric with respect to the maximum. In particular, the  right (left)
tail of the distribution of $\mathcal{Q}^{(2)}$ ($\mathcal{Q}^{(1)}$) is  visible fatter than
the left (right) one. As a further test, in the inset of the same figure we tried to collapse
the curves similarly to what done in the upper panel for the case above $T_c$, but, due to the
non Gaussian behavior, the collapse fails. In order to study the shape of the PDs more
carefully, in Fig.~\ref{pdfmendezaboveTcbis} we plot the logarithm of the PD for $Q^{(1)}$
(normalized by the height of $A_\tau$ of the peak) against $|Q^{(1)}(\tau)-\langle
Q^{(1)}(\tau)\rangle |/\sigma _\tau $, where $\sigma^2 _\tau $ is the variance
 of the distribution proportional to $\tau$. In this representation, a Gaussian PD should
correspond to a power law, namely a straight line of slope $2$. The figure shows that the
Gaussian behavior is approached as $\tau $ increases. For finite $\tau $, however, the
distribution is strongly non Gaussian both around the average $Q^{(1)}\simeq \langle
Q^{(1)}\rangle$ and on the tails. In particular, the slope of the {\it positive } tail with
$Q^{(1)}\gg \langle Q^{(1)}\rangle$ (lower panel) is always smaller than that of the {\it
negative } one, confirming that the former is fatter. In addition, the approach to a Gaussian
behavior is faster for the negative tail. This pattern of behavior can be expressed
quantitatively by the skewness $ S(\tau )=<(\mathcal{Q}(\tau) - <\mathcal{Q}(\tau)>)^3>/
<(\mathcal{Q}(\tau) - <\mathcal{Q}(\tau)>)^2>^{3/2}$ of the distributions. This quantity is
plotted in the inset of Fig.~\ref{pdfmendezaboveTcbis}, showing a power law decay consistent
with $S(\tau )\sim \tau ^{-a}$, with $a \simeq 1/2$.

Regarding the slopes $D^{(n)}(\tau)$, they converge to 1 and the scaling $\epsilon(\tau,L)
\sim 1/x$ with  $x=\tau$ is verified, as shown in Fig.~\ref{scalingmendez} (we have not
reported  the data for $\mathcal{Q}^{(2)}$ which are more noisy than those for
$\mathcal{Q}^{(1)}$). We  observe that a fit based on Eq.~(\ref{deviazione1}) well reproduces
the simulation data also in this case.  As anticipated, the times required to reach the slope
1 are much longer than the corresponding ones at high temperature (but always smaller than
$\tau _{erg}(N,T_2)  \gtrsim 5 \times 10^8$ MCS). These results suggest that the
FR~(\ref{FR2T}) and the scalings (\ref{deviazione1},\ref{scaling}) hold even in states with
symmetry breaking and that the presence of the macroscopic timescales $\tau _{erg}(N,T_n)$ do
not affect their validity. Recalling the discussion at the beginning of this section, it is
reasonable to conjecture that for  $\tau \ll \tau _{erg}(N,T_2)$ the system enters a NESS
which is representative of the state  with broken symmetry that one would have (indefinitely)
for an infinite system. For the same mechanism discussed above in  equilibrium conditions, in
this state there is a single relaxation time left (playing the role of $\tau _{eq}(T)$), and
the FR is obeyed on timescales larger then this microscopic time. Moreover, recalling the
discussion of Sec.~\ref{intro}, this case is particularly interesting since the FR is verified
(although with the preasymptotic corrections described by Eq.~(\ref{deviazione1})) in a far
from equilibrium situation in which the PDs are not Gaussian.

\section{Stationary states under shear}

In the present section we consider a system which is coupled to a single heat bath and is
mechanically driven out of equilibrium. Mechanical work is done on the system by shearing it
in the horizontal direction. Boundary conditions are periodic in the horizontal direction
(parallel to the shear) and open in the vertical direction. In a shear event, the horizontal
line with coordinate $y \in \{1, \cdots L\}$  is moved by $ \lambda y$ lattice steps to the
right. Shear events occur at regular intervals of $r$ elementary Monte Carlo moves, with the
shear period $r$ submultiple of $N = M \times L$. For each Monte Carlo sweep (MCS) over the
lattice there are $N/2$ Kawasaki moves and $N/r$ shear events. This model can be thought of as
a discrete version for the kinetic equation of a binary mixture subject to the convective
velocity $v_x(y) = \lambda y/(r/N)$ having a shear rate $\dot\gamma = \frac{dv_x}{dy} =
\lambda N / r$. The  model, with single--spin--flip dynamics, was used to study domain growth
properties of sheared systems in \cite{CGS}.

The heat released to the bath during the evolution will be computed as in Eq.~(\ref{calore}),
where the index $(n)$ will be dropped since here the system is exchanging energy with a single
reservoir at temperature $T$. Work done on the system is instead defined as the energy
variation in shear events.

 Starting from a random configuration, with shear applied at a
constant rate, a steady state is reached. As before, the PDs for heat and work relative to the
steady state can be constructed collecting the values integrated over segments of length
$\tau$ in a long trajectory.

In the following we report results obtained for a $50 \times 4$ lattice, with $T=5$, shear
step $\lambda = 1$ and shear period $r = 50$, which corresponds to a shear rate $\dot\gamma =
4$. Data have been taken mainly with $10^7$ MCS to reach the NESS and $10^8$ MCS for the
measurements. A few longer ($5 \cdot 10^8$ MCS) runs have been made for some data points.

\begin{figure}
\hskip 2cm
\includegraphics*[height=6.5cm,width=11.cm]{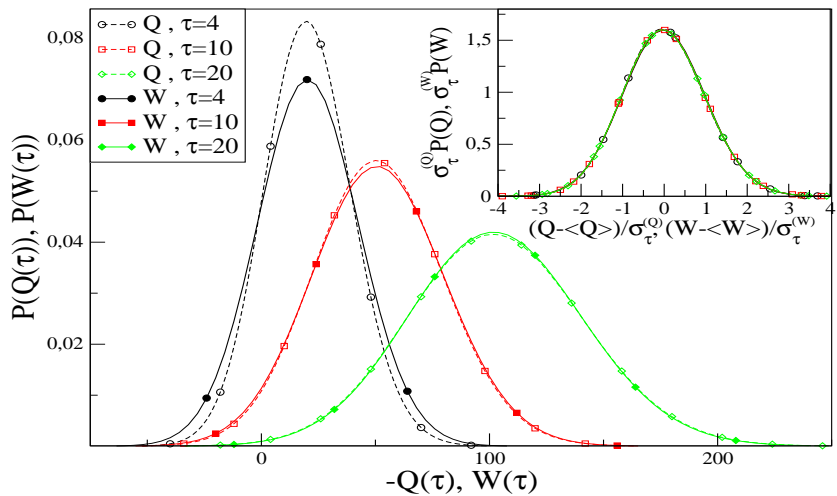} \vskip -1.4cm
\vskip 1cm \caption{Heat and work PDs for the system under shear at different $\tau$. $5 \cdot
10^7$ MCS have been
  discarded in order to reach stationarity and $5 \cdot 10^8$ MCS have
  been used for measurements. See text for the other parameters.  In the inset
  the curves are separately rescaled as in Fig.~\ref{pdf_kawas_aboveTc}
  ($\sigma _\tau ^{(Q)}$ and $\sigma _\tau ^{(W)}$ are the standard deviations
  of $\mathcal{Q}$ and $W$, respectively, for different $\tau$).}
\label{Shear-PDs}
\end{figure}

In Fig.~\ref{Shear-PDs}  we show the work and heat PDs.  At smaller $\tau$ the heat
distributions are more peaked than the work PDs while they become more similar when $\tau$
increases.  In all cases the PDs are very well fitted by Gaussian distributions and the work
and heat PDs at different $\tau$ collapse on a master curve when rescaled as in
Fig.~\ref{pdf_kawas_aboveTc}.

\begin{figure}
\hskip 2cm
\includegraphics*[height=6.5cm,width=11.cm]{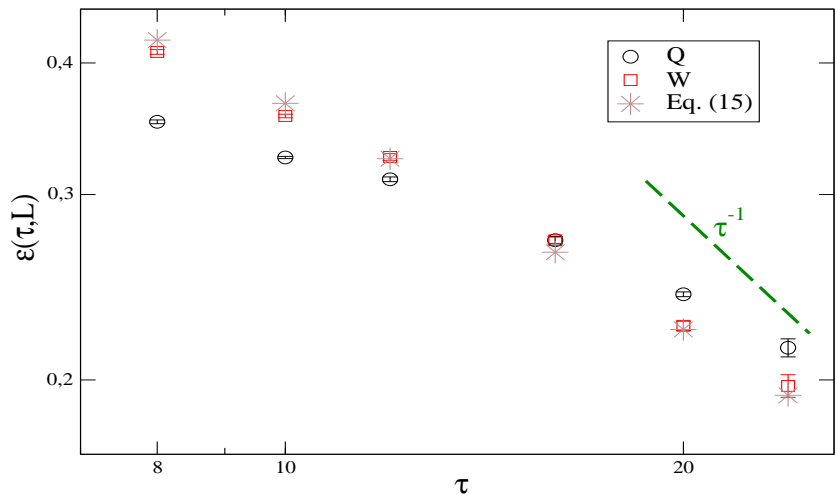} \vskip -1.4cm
\vskip 1cm \caption{ $\epsilon^{(\mathcal{O})}$ ($\mathcal{O}= \mathcal{Q},\mathcal{W}$) as a
function of $\tau$ for the same system of Fig.~\ref{Shear-PDs}.} \label{Shear-Slopes}
\end{figure}

As in Sect. 2,  also here we define  the distance from the expected asymptotic behavior
$\epsilon^{(\mathcal{O})}=1-D^{(\mathcal{O})}(\tau)$ where, analogously to Eq.~(\ref{slope}),
the slope $D^{(\mathcal{O})}(\tau)$ is defined as
\begin{equation}
D^{(\mathcal{O})}(\tau) = \pm
 \frac {\ln{\left({\mathcal{P}(\mathcal{O}(\tau))}/
{\mathcal{P}(-\mathcal{O}(\tau))}\right)}} { \mathcal{O}(\tau)/T}
 \label{slopeshear}
\end{equation}
where the  signs $+,-$ are for $\mathcal{O}=\mathcal{W},\mathcal{Q}$, respectively. In
Fig.~\ref{Shear-Slopes} we report the behavior of $\epsilon^{(\mathcal{O})}$ as a function of
$\tau$. $\epsilon^{(\mathcal{O})}$ tend to zero, but differently than in the case of
thermodynamic driving, the approach to the expected asymptotic result is slower, especially
for the heat.

 Analogously to the case of two thermostats, the ratio of the probability of a trajectory
over that of the time-reversed is given by
\begin{equation}
\frac{\mathcal{P} (traj)}{\mathcal{P}(-traj)} = e^{-\frac {\mathcal{Q}(\tau)}{T} } =
e^{\frac{\mathcal{W}(\tau)}{T} - \frac{\Delta E}{T}}  , \label{microrevshear}
\end{equation}
where, for each trajectory, the relation $\mathcal{Q}(\tau) + \mathcal{W}(\tau) = \Delta E$
holds. Proceeding as in the case with thermodynamic drive one arrives at an expression similar
to Eq.~(\ref{deviazione1})
\begin{equation}
{\epsilon^{(\mathcal{O})}} = \mp \frac{T \langle \mathcal{O}(\tau)
\rangle}{(\sigma^{(\mathcal{O})}_{\tau})^2} +  T \sqrt{\left( \frac{\langle \mathcal{O}(\tau)
\rangle}{(\sigma^{(\mathcal{O})}_{\tau})^2}\right) ^{2}+\frac{v^2}
{(\sigma^{(\mathcal{O})}_{\tau})^2} },
 \label{deviazioneshear}
\end{equation}
where the signs $-,+$ are for $\mathcal{W},\mathcal{Q}$ respectively, and  $v^2 $ is an
unknown quantity that can be used as a fit parameter. In doing that we find a very good
agreement between formula (\ref{deviazioneshear}) and the numerical data for $W$, while the
accordance (not shown in Fig.~\ref{Shear-Slopes}) is very poor for $\mathcal{Q}$, probably
because, as already noticed, this quantity has a slower convergence. The faster convergence of
$W$ can be possibly
 ascribed to the fact that, after making the quasi-equilibrium assumption
$\frac{\mathcal{P}^{staz}(\tau)}{\mathcal{P}^{staz}(0)} \simeq e^{-\beta \Delta E} $ and
calculating the averages over trajectories as in Sec. \ref{finitetau}, the boundary term
proportional to $\Delta E$ cancels for the work while it remains when one consider the
behavior of the heat.

\section{Discussion and Conclusions} \label{concl}

In this Article we have considered the issue of the FRs in the Ising model with nearest
neighbor interactions. We focused on two different NESS, induced by a thermodynamic drive
provided by two thermostats at different temperatures, or by a mechanical forcing realized by
shearing the system. We find that the FR (\ref{FR}) is verified in any case when $\tau $ is
sufficiently large. For the case with different thermostats, we have also provided an exacting
test of the FR in the far from equilibrium regime where  deviations from the linear response
theory are observed with  the Green-Kubo relation
 violated (above $T_c$), or the PDs are not Gaussian (below $T_c$). Furthermore, we have
analyzed  the effects of finite time by proposing a scaling law which takes into account the
geometry of the system and the details of the interaction with the baths and that is well
verified numerically. According to this law the leading corrections to the FR (\ref{FR}) decay
as $\tau ^{-1}$, in agreement with what found in some previous studies \cite{GC95,vZCC,cili1},
while the dependence on the system size relies on the details of the coupling to the baths.
The derivation of our scaling law is based on few strong but general hypotheses, which could,
in principle, be verified also in different systems. A natural perspective for future work,
therefore, is the understanding of the possible generality of our results in different
contexts.

Most of the results of this paper are obtained for systems that are able to equilibrate in the
small entropy production limit \cite{drivenmean}. Basically this means that, for this class of
systems, the equilibrium state is recovered in a finite time when the drive is switched off.
In this case the FR generally holds, as we have explicitly verified. The generalization of the
FR to systems which do not equilibrate in the small entropy production limit is a very
interesting issue. Non-equilibration typically happens when the limit of large $\tau $ is
taken after the thermodynamic limit in systems, such as glasses or ferromagnets below the
critical temperature, whose equilibration time diverges with the system size. Regarding these
non-equilibrating systems, some conjectures have recently appeared in the literature \cite
{sellitto}, inferring that a relation similar to (\ref{FR}) holds, where, however, the entropy
production, instead of being a function of the bath(s) temperature (as for instance in
Eqs.(\ref{FR2T},\ref{FRW})), depends on $T_{eff}$, namely the effective temperature
\cite{teff} which characterizes the dynamics of these systems. For the model considered in
this paper one has $T_{eff}=\infty$ \cite{tinfty}. In this perspective, therefore, it would be
very interesting to study the technically demanding case of the Ising model with shear for
$T<T_c$.

\ack
 The authors are grateful to G. Saracco and L. Rondoni    for useful discussions.

\end{document}